# Wide-band optical field concentrator for low-index core propagation

**D. de Ceglia, M. De Sario, A. D'Orazio, V. Petruzzelli, M.A. Vincenti**

*Dipartimento di Elettrotecnica ed Elettronica, Politecnico di Bari*

*Via Orabona 4, 70125 Bari, Italy*

**F. Prudenzano**

*Dipartimento di Ingegneria dell'Ambiente e per lo Sviluppo Sostenibile*

*Politecnico di Bari - II Facoltà di Ingegneria*

*Viale del Turismo 8, 74100 Taranto, Italy*

**M. Scalora**

*Charles M. Bowden Research Center, AMSRD-AMR-WS-ST,*

*Redstone Arsenal, AL 35898-5000, USA*

We propose a novel chirped structure consisting of a low index polymer core bounded by modulated multilayer claddings, to realize an optical field concentrator with virtually zero propagation losses in a wide spectral range, independent of wave polarization. In spite of the absence of the total internal reflection mechanism, properly designed multilayer claddings ensure the achievement of unitary transmittance in a wide spectral range, including the widely used wavelengths for optical communications. Several cladding geometries obtained by varying the thicknesses of the cladding layers are reported and discussed.
**Keywords:** photonic crystal, polymer, Bragg reflection waveguide

# 1 INTRODUCTION

High-index cores sandwiched by low index claddings are used for conventional optical field confinement. In fact, guided modes require total internal reflection (TIR) to achieve optical field concentration and propagation. Another way to confine the light flow is offered by the use of periodic structures. Light within certain wavelength ranges and wave vector directions is not allowed to propagate through these periodic structures, giving rise to photonic band gaps (PBGs) [1]. The well known ability of controlling light using photonic crystals (PCs) has been exploited to demonstrate a wide range of integrated optics applications. For example, using the external reflection due to dielectric layer interferences, a simple PC, such as a one-dimensional arrangement of two different dielectrics, can be easily used as an optical filter. Another intriguing property of these artificial materials is the strong, anomalous, geometrical dispersion derived by their intrinsic periodicity. Such a dispersion, which essentially depends on the high index contrast [2], could be much greater than the normal chromatic dispersion of many media, leading to devices capable of enhanced efficiencies in the framework of nonlinear optics [3,4], such as second harmonic generation [5,6]. If the periodicity of the structure is broken by introducing one or more defects, localized states appear inside the band gap or gaps: light becomes confined in small regions of space, reaching high density of modes and exceptionally low group velocity. This property may be exploited in the design of slow light structures [7] and microcavities [1]. Moreover, the Bragg confinement of light provided by photonic crystals is a suitable tool for light propagation, representing a valid alternative to the total internal reflection (TIR) effect. Hence, we are able to divide light guidance mechanism in two different categories:

traditional high-index core waveguides based on the TIR mechanism and photonic band gap waveguides in which light propagates through the low-index core due to the Bragg scattering mechanism. Different kinds of devices are based on Bragg scattering, such as Bragg reflection waveguides (BRWs) [8], Bragg fibers [9] or photonic crystal fibers (PCFs) [10,11]. Different geometries of PBG fibers, built up by concentric rings [12] of alternating high-index and low-index layers surrounding a low-index core, were theoretically studied [13,14] and experimentally demonstrated [15]. The fundamental issues of the coaxial omni-guide structures are the possibility to overcome problems of traditional TIR-based waveguides, such as polarization rotation, pulse broadening or sharp bending. Easy coupling, broadband transmission together with optimal field confinement, tuning of single-mode window and frequency of zero dispersion are some of the exciting features of waveguides based on the Bragg scattering mechanism [13,14]. Photonic-crystal-assisted waveguides, having a low index core bounded by properly designed multilayer stacks, have been recently proposed as a coaxial waveguide [16] or as a high-performance Mach-Zehnder modulator [17] and harmonic generator [18].

In this paper we propose an optical field concentrator operating as a waveguide in a much wider spectral range. Firstly, to show the ability of confining light in a low-index core, a simple multilayer periodic cladding is studied and discussed. Secondly, we suggest a low-index polymeric core layer bounded by an aperiodic, chirped multilayer cladding, dimensioned by laws which guarantee a uniform distribution of layer thicknesses varying between a minimum value $L_{min}$ and a maximum value $L_{max}$. The latter structure achieves a unitary transmittance in a wide range of wavelengths, including all the three optical-fibers communication windows. Furthermore, the device acts as a high-field concentrator in the low index region with virtually zero propagation losses, provided material losses due to linear absorption and scattering phenomena induced by fabrication defects may be neglected. Finally, we show that guided modes have effective refractive indices smaller than the refractive index of each of the cladding layers.

## 2 LOW-INDEX CORE WAVEGUIDING ASSISTED BY 1-D PHOTONIC CRYSTAL CLADDING

A symmetric geometry constituted by a 2 μm thick core having refractive index $n_c$=1.6 (typical for polymers) bounded by a periodic stack-cladding, as sketched in Fig.1, is first studied. Each period is composed by two alternating layers, having refractive indices $n_1$=$n_c$ and $n_2$=3.2 (typical for $Al_xGa_{1-x}As$ with x=0.6 for λ=1064 nm), and the same thickness d. This choice is by no means unique and it leads to a simpler way to design the transverse profile of the structures. The Beam Propagation Method (BPM), in combination with an analytical approach based on mode-matching at each interface, is used to analyze these structures of finite transverse and longitudinal extent. Measured values of material dispersion are also considered in all the simulations [19,20]. The waveguides are excited with a cosine field profile, corresponding to the fundamental TE-mode shape, placed in x=0 and propagating along z. The input signal propagates down a one-millimeter-long structure. The longitudinal transmittance is shown in Fig.2 for claddings having fixed layers of 150 nm, 180 nm, 210 nm, and 240 nm, respectively. Propagation is allowed only in certain bands, as a consequence of the Bragg-mirror behaviour of the lateral multilayer stacks. By varying layer thicknesses, the transmittance band shifts towards higher wavelengths [17]. If the signal propagates through the transverse (x axis) direction, the structure shows high-localized resonant states at frequencies which fall inside the band gap, acting as a defective, one-dimensional photonic crystal. For example, using 150nm thick layers, two resonant modes appear at λ=1340 nm and λ=1590 nm.

# 3 LOW-INDEX CORE ASSISTED BY A CHIRPED MULTILAYER CLADDING

In this section we will show the effect of the presence of chirped claddings on the transmission properties. As shown in Fig.2, the transmission bands red-shift by increasing the thickness of each layer, i.e. the period of the cladding. We exploit this interference effect to cover the band gap regions in the transmission spectrum in order to realize a wide-band device [21], "including" more than one period thickness in the cladding. This can be achieved by modulating the period thickness along the transverse direction between a minimum value and a maximum one, which in all the simulations have been assumed equal to $L_{min}$ = 120 nm and $L_{max}$ = 240 nm. We start our analysis by considering a linear modulation of the type:

$$d_n = L_{min} + n * \Delta \qquad (1)$$

where $\Delta$ is the difference between two adjacent layers, $d_n$ is the single layer thickness and n = 0,1,2…N, so that the first layer nearby the core is the thinnest and the last layer is the thickest. N+1 is the number of the periods constituting the cladding. Each period is composed by two layers. Once the first and the last layers of the periods have been fixed after observing the band pass displacement, the value of $\Delta$ is determined by the choice of N. The resulting transmission spectra versus wavelength calculated by using values of $\Delta$ between 8 nm and 20 nm are plotted in Fig.3. Each spectrum is calculated using a fundamental TE-mode-shaped input that propagates one millimeter along the longitudinal direction (z axis). The transmission coefficient is almost unitary, reaching in the worst case ($\Delta$ =20 nm) the value 0.9985 for $\lambda$ = 1600 nm. To properly weigh the optical field confinement by varying the $\Delta$ value, we introduce the figure of merit (FOM) C defined as follows:

$$C = \int_\lambda \frac{\int_{core} E^2 dx}{\int_{total} E^2 dx} \, d\lambda , \qquad \lambda = 400,\ldots 1600 \text{ nm} \qquad (2)$$

where E is the electric field evaluated after one millimeter of propagation and C is obtained by averaging out in the operating frequency range. By changing the N value we obtain a trade-off between a wide-band, unitary transmittance, and good field confinement inside the core region, as shown in Tab.1:

| N | T [µm] | $\Delta$ VALUE [nm] | C |
|---|---|---|---|
| 16 | 13.52 | 8 | 0.9171 |
| 13 | 11.36 | 10 | 0.9237 |
| 9 | 8.48 | 15 | 0.9438 |
| 7 | 7.04 | 20 | 0.9323 |

Tab.1 : Optical field confinement values

where N is the total number of periods and T is the transverse length of the device. The maximum value of C is reached for $\Delta = 15$ nm.

The field profiles for the linearly modulated cladding ($\Delta = 15$ nm) at $\lambda=1064$ nm (dashed-dot line), $\lambda=1320$ nm (solid line) and $\lambda=1550$ nm (dashed line) are shown in Fig.4. Losses are exceptionally small: 0.009 dB/cm for $\lambda=1064$ nm and 0.004 dB/cm for $\lambda=1320$ nm and $\lambda=1550$ nm.

If the beam propagates along the x-axis, a linearly modulated structure is seen as a multiple-couple-cavity system, giving several resonant modes across the entire spectrum range under consideration ($\lambda = 400 \div 1600$ nm).

By imposing the continuity of the fields at each interface, we evaluate the effective refractive indices of all the modes supported by the structure. The multilayer claddings operate as mirrors for the light inside the core region, because of the Bragg nature of the multilayer structure, thus yielding an effective refractive index of guided modes smaller than the index in each of the cladding layers [2-3]. In fact, the effective refractive index of the first mode is $n_{eff}=1.5777$ for $\lambda=1064$ nm, $n_{eff}=1.5615$ for $\lambda=1320$ nm and $n_{eff}=1.4993$ for $\lambda=1550$ nm, provided no material dispersion is considered. The linearly chirped optical concentrator and the others described below are generically able to support multimode propagation. The key design parameter to achieve monomodality is the core width. Moreover, by analyzing the structure with a finite-element method based commercial software, we verified that this device has the same behaviour for TE and TM polarization (Fig.5).

Another interesting property of these structures is the relation between the unitary transmittance and the period thickness instead of their spatial position into the cladding. To underline this peculiarity, a reverse linearly modulated cladding was considered by using relation (1), but in this case the thicker layers are in the neighboring of the core.

The transmittance curve for this cladding structure highlights that there are no meaningful differences between the linearly modulated and reverse linearly modulated claddings in terms of propagation losses and transmission spectra; we notice also how the lossless feature is retained reversing the cladding.

## 4  ALTERNATIVE CHIRPING LAWS

To stress the flexibility of chirped, modulated claddings, various modulation laws were tested, following the relations listed in Tab.2.

| **CHIRPING LAW** | **LAYER EQUATION** |
|---|---|
| EXPONENTIAL LAW | $d_n = d_{n-1} + \partial^{\,n}, \quad n = 0,1,2…N$ |
| RANDOM LAW | $d_n = L_{min} + random * L_{min}, \quad n = 0,1,2…N$ |
| BINOMIAL LAW | $d_n = L_{min} + \dfrac{c}{2} * \dfrac{L_{max}}{\max(c)}, \quad c = 1,4,6,4,1; \quad n = 0,1,2…N$ |
| GAUSSIAN LAW | $d_n = L_{min} + gauss(i) * L_{min}, \quad n = 0,1,2…N$ |

Tab.2: Chirping laws

The cladding governed by the exponential chirping law is constituted by nine bi-layers periods, where $d_0 = L_{min}$, $\partial = 1.5$ nm, n is the order number of the period, as in the linear chirping law studied earlier, whose thickness increases starting from the core of the cladding.

The transverse length of this structure is T = 7.5 µm (less than the linearly and reverse-modulated claddings).

Both the linear and exponential modulations of the cladding show the same behavior regardless of the spatial position of the periods: a uniform distribution of layer thickness varying between $L_{min}$ and $L_{max}$ within the cladding ensures zero propagation losses in the spectral range of our interest. To highlight the quality of the structure having uniform distribution of layer thickness, a random law was tested (Tab.2). We considered nine periods where n is the period index and the *random* function gives back a uniformly random value between 0 and 1. The total width of the structure depends now on the *random* function. In the case of the binomial chirped structure, the binomial law was obtained by tailoring a $4^{th}$ order binomial, where c is the binomial coefficient. The n subscript is the period index and the max(c) function gives back the maximum binomial coefficient, that is 6. The cladding is built up with five periods, composed of two layers. The total thickness of the structure is T = 5.7 µm, less than that of the aforementioned structures. In this case we do not achieve a unitary transmittance in the frequency range of our interest having a non uniform period thickness distribution (between $L_{min}$ and $L_{max}$) .

Finally, to achieve the best performance with symmetric modulated claddings the Gaussian law was considered. The function *gauss(i)* gets 9 values carried out by considering a uniform distribution in the range [-5,5], with a Gaussian function given by µ=0 and σ = $1/(\sqrt{2\pi})$. The total thickness of this device is T = 7.9 µm. As achieved in the other configurations, losses are indeed very small: 0.013 dB/cm for λ=1064 nm, 0.008 dB/cm for λ=1320 nm and 0.0001 dB/cm for λ=1550 nm. As expected, good field confinement is achieved also for the last two different modulated claddings. For the sake of clarity, the transmittance spectra of all the structures presented in the paper are shown in Fig.6.

## 5 CONCLUSION

A strong field confinement in low refractive index media, such as polymers, is reached if guiding inside the low-index core is assisted by multilayer claddings. The modulation of the multilayer geometry allows us to retain good field confinement in a wide spectral range including all the three optical-fiber communication windows. The performance of such devices in terms of transmission coefficient over a wide spectral range strictly depends on the period thickness rather than the spatial distribution of each period within the cladding. The confinement properties of these structures, including higher order mode propagation, losses and their effective refractive indices were discussed as well.

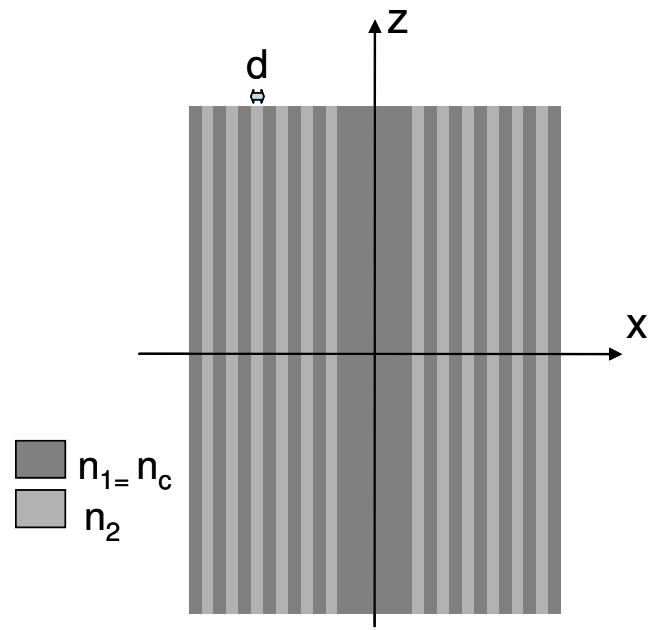

Fig.1: Photonic-crystal assisted waveguide. Light grey regions have $n_1=n_c=1.6$ and dark grey regions have $n_2=3.2$ and d indicates the single layer thickness

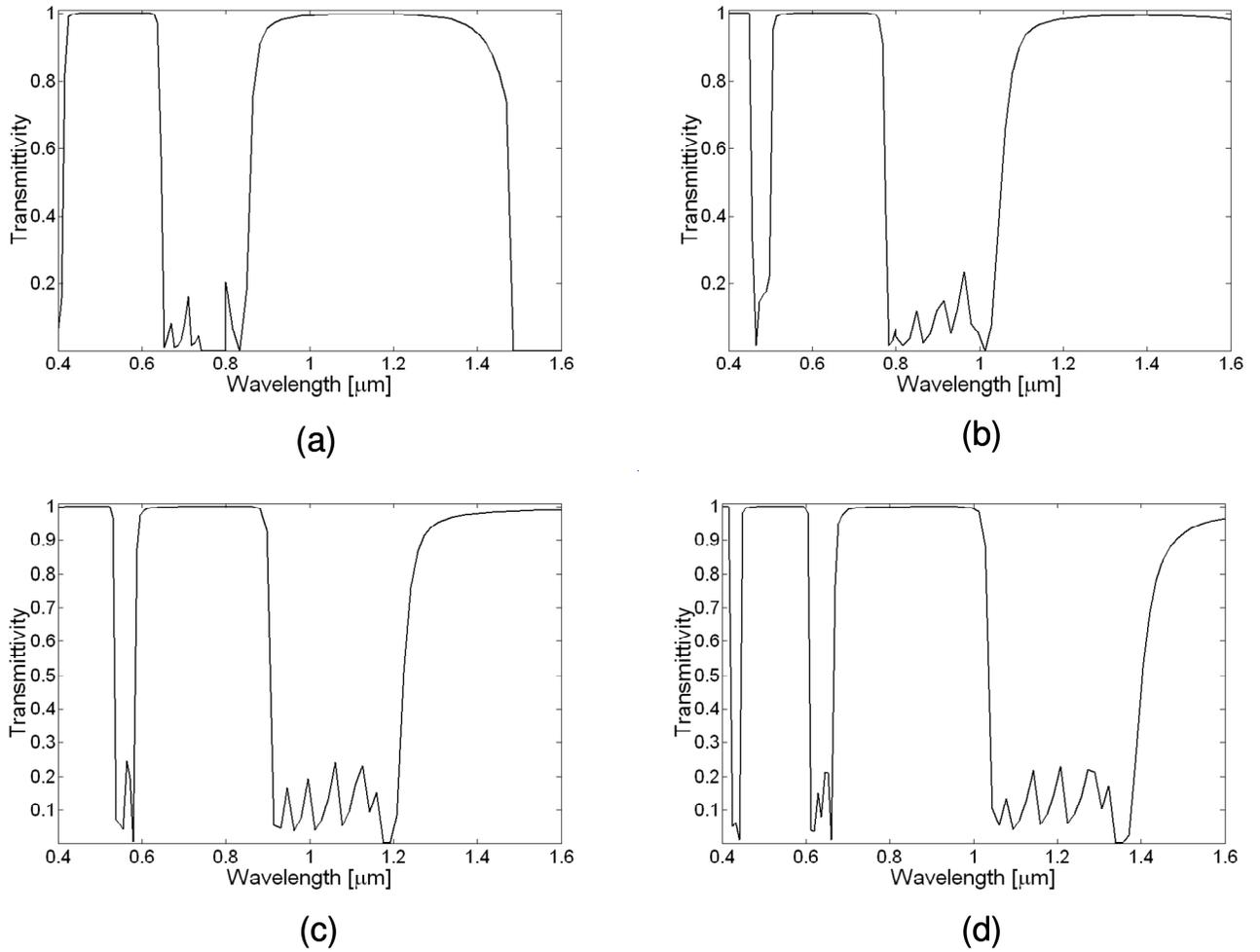

Fig.2: Transmittivity of fixed period waveguides: (a) 150 nm fixed layers cladding; (b)180 nm fixed layers cladding; (c) 210 nm fixed layers cladding; (d) 240 nm fixed layers cladding

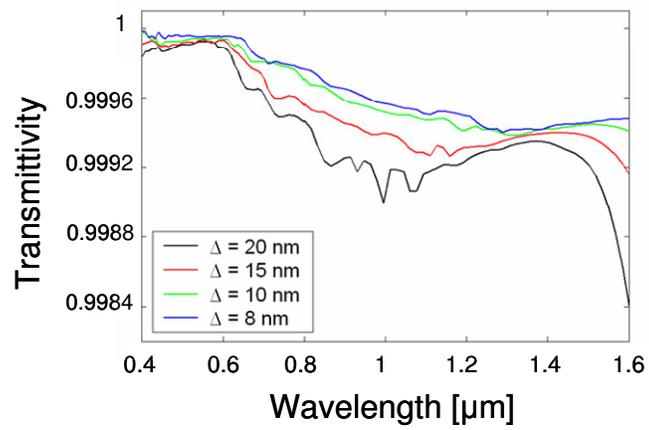

Fig.3: Transmittivity dependence on the Δ value of the optical concentrator having linear chirped cladding

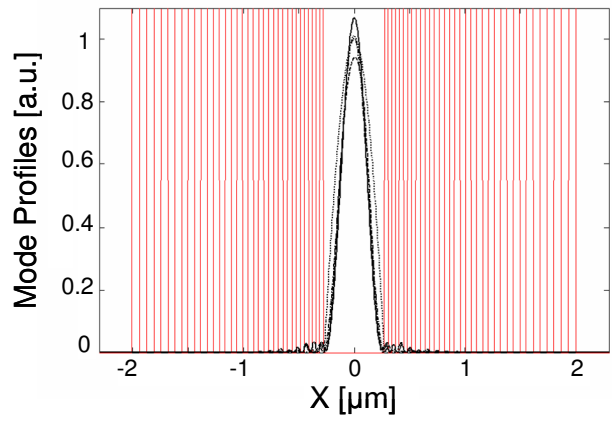

Fig.4: Profiles of the mode intensities at λ=1064 nm (dashed-dot line), λ=1320 nm (solid line) and λ=1550 nm (dashed line) for linear modulated cladding device. Dotted line is for input mode profile

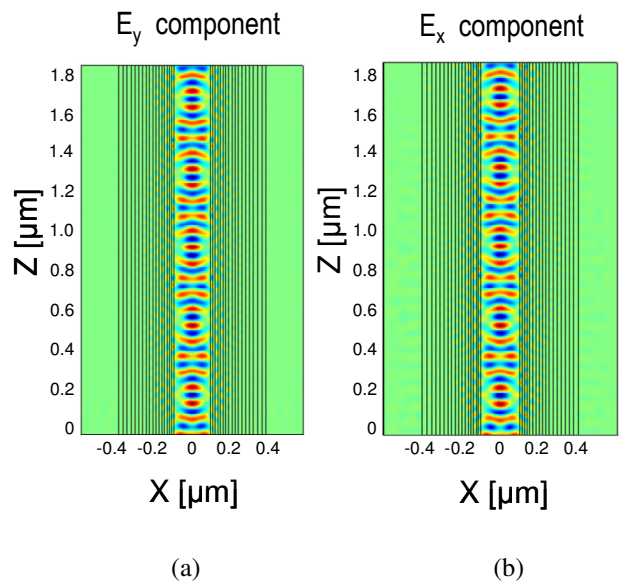

Fig.5: Field localization for TE (a) and TM (b) polarization modes @ λ=1320 nm

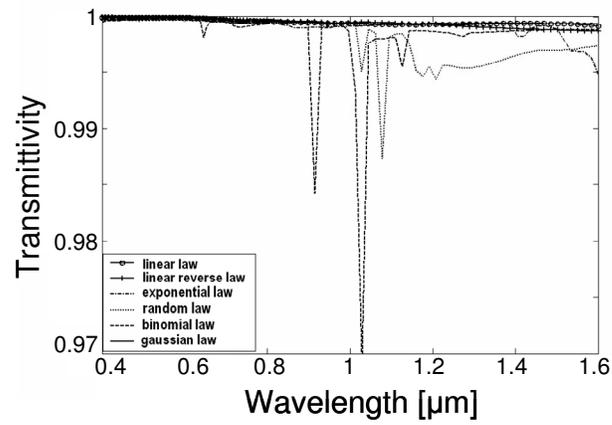

Fig.6: Transmittivity for linear modulated cladding waveguide (solid line with circle markers), reverse linear modulated cladding waveguide (solid line with plus markers), exponential modulated cladding waveguide (dashed-dot line), random modulated cladding waveguide (dotted line), binomial modulated cladding waveguide (dashed line) and Gaussian modulated cladding waveguide (solid line)